# Comparative Analysis of Deep Learning Models for Olive Tree Crown and Shadow Segmentation Towards Biovolume Estimation

Wondimagegn Abebe Demissie
*Institute of Crop Science*
*Scuola Superiore Sant'Anna*
Pisa, Italy
wondimagegn.demissie@santannapisa.it

Stefano Roccella
*Biorobotics Institute*
*Scuola Superiore Sant'Anna*
Pisa, Italy
stefano.roccella@santannapisa.it

Rudy Rossetto
*Institute of Crop Science*
*Scuola Superiore Sant'Anna*
Pisa, Italy
rudy.rossetto@santannapisa.it

Antonio Minnocci
*Institute of Crop Science*
*Scuola Superiore Sant'Anna*
Pisa, Italy
antonio.minnocci@santannapisa.it

Andrea Vannini
*Biorobotics Institute*
*Scuola Superiore Sant'Anna*
Pisa, Italy
andrea.vannini@santannapisa.it

Luca Sebastiani
*Institute of Crop Science*
*Scuola Superiore Sant'Anna*
Pisa, Italy
luca.sebastiani@santannapisa.it

*Abstract*—Olive tree biovolume estimation is a key task in precision agriculture, supporting yield prediction and resource management, especially in Mediterranean regions severely impacted by climate-induced stress. This study presents a comparative analysis of three deep learning models U-Net, YOLOv11m-seg, and Mask R-CNN for segmenting olive tree crowns and their shadows in ultra-high resolution UAV imagery. The UAV dataset, acquired over Vicopisano, Italy, includes manually annotated crown and shadow masks. Building on these annotations, the methodology emphasizes spatial feature extraction and robust segmentation; per-tree biovolume is then estimated by combining crown projected area with shadow-derived height using solar geometry. In testing, Mask R-CNN achieved the best overall accuracy (F1 ≈ 0.86; mIoU ≈ 0.72), while YOLOv11m-seg provided the fastest throughput (≈ 0.12 second per image). The estimated biovolumes spanned form ~ 4 - 24 m³, reflecting clear structural differences among trees. These results indicate Mask R-CNN is preferable when biovolume accuracy is paramount, whereas YOLOv11m-seg suits large-area deployments where speed is critical; U-Net remains a lightweight, high-sensitivity option. The framework enables accurate, scalable orchard monitoring and can be further strengthened with DEM/DSM integration and field calibration for operational decision support.

*Keywords—Instance segmentation, deep learning, Unmanned Aerial Vehicle, olive cultivation, biovolume estimation*

## I. Introduction

Olive (*Olea europaea* L) tree cultivation is a vital agro-industrial activity that significantly contributes to the economic and social development of countries within the Mediterranean agroecosystem [1]. According to data from the International Olive Council (IOC), over 95% of olive tree cultivation is in the Mediterranean basin [2], generating over €7 billion in production every year for southern European countries [3], [4].

However, prolonged drought and heatwave have severely affected the olive yields across major producing countries, with Spain and Greece experiencing production losses of up to 50% [5]. These extreme climatic conditions place considerable stress on olive trees, leading to premature fruit drop and a decline in both oil quality and yield [5]. To mitigate such challenges modern agricultural mitigation and adaptation strategies are required.

A critical task in modern agricultural practices particularly in precision agriculture is the automated estimation of olive tree crowns using high resolution Earth observation imagery, which represents the fundamental step in assessing and monitoring both the productivity and health of olive trees [6]. Precision agriculture has undergone transformative advancements through the integration of UAV-based remote sensing, which enables high-resolution monitoring of crop health, growth patterns, and yield potential at the field to medium scale of agriculture [7], [8]. Advanced UAV platforms have become a popular technique for identifying and counting trees and determining height and diameter, crown area and biovolume of trees [6], [10]. According to [8] and [11], hyperspectral and multispectral imaging technologies mounted on UAV provide high-resolution, field-scale agricultural information by capturing data across multiple spectral bands.

Despite advances, extracting information from UAV crop images remains challenging due to complex field condition (varying light, soil reflectance, and weeds resulting in alterations in color, texture and shape of UAV images) and diverse image capture factors (shooting angles and camera resolution resulting in potential blurring of UAV images) [10].

Recently, various deep learning (DL) algorithms have shown great promise in overcoming such challenges by effectively extracting structural information and reducing the high dimensionality of large UAV-based datasets [10]. Convolutional neural networks (CNNs) have significantly

The work described in the present paper has been developed within the project entitled VIRMA: Velivolo Intelligente Robotico per il Monitoraggio Agro-Ambientale - funded by the MUR within the framework of the Fund for the promotion and development of policies of the National Research Program (PNR), in accordance with EU Regulation no. 241/2021 and with the PNRR 2021-2026. The sole responsibility of the issues treated in the present paper lies with the authors.

outperformed over conventional image classification methods such as random forest, extreme gradient boosting and support vector machine by effectively capturing spatial features from UAV images [6], [12]. CNNs models have consistently demonstrated strong performance across a wide range of computer vision tasks, including image classification [13], object detection [14], semantic segmentation [15], [16] and instance segmentation [17].

Semantic and instance segmentation models are increasingly utilized for their ability to detect and delineate group and individual objects within images. This capability is crucial for advancing precision agriculture applications, such as tree crown delineation, individual-level tree monitoring and biovolume estimation.

This paper presents a comparative analysis of deep learning-based segmentation models for estimating the biovolume of olive trees using ultra-high-resolution UAV imagery. The approach leverages spatial features derived from multispectral UAV images to segment crown areas and corresponding heights (from shadow lengths) to estimate biovolume. Through comparative analysis and experimental validation, we intend to demonstrate the added value of integrating advanced deep learning techniques and UAV data fusion for precise object-level segmentation. The objective of this study is to illustrate how these technologies can enhance the accuracy, scalability, and automation of biovolume estimation in precision agriculture. This contributes to more informed decision-making in crop monitoring, resource allocation, and yield forecasting, particularly in the context of climate-induced agricultural stress.

## II. RELATED WORKS

The application of deep learning to UAV-based tree monitoring has attracted considerable attention, particularly for olive and orchard systems where crown delineation and biovolume estimation are essential. Mask R-CNN has consistently demonstrated strong performance in instance segmentation, enabling precise separation of overlapping crowns and the integration of crown and shadow information for volumetric estimation. [6] reported accuracies above 80% in olive biovolume prediction, underscoring its potential for object-level monitoring.Similarly, [18] employed Mask R-CNN on UAV RGB and multispectral data (NDVI, GNDVI), reporting crown segmentation F1-scores of 95–98% and demonstrating that integrating spectral indices improved biovolume estimates (about 82% accuracy against field-measured volumes). These works confirm that accurate crown area extraction together with shadow-based height can serve as a reliable proxy for olive tree volume and biomass. Yet, Mask R-CNN's two-stage architecture results in high computational demand, limiting its suitability for large-scale or real-time deployment.

Alternatively, U-Net provides efficient semantic segmentation with an encoder–decoder structure and skip connections that preserve spatial detail. It has been successfully applied in orchard mapping, though its class-level masks restrict per-tree delineation in dense plantations. Recent enhancements, such as the ECA-U-Net with channel attention, improved mIoU (49.19%) and overall accuracy (85.87%) compared with standard U-Net, particularly in edge preservation and small-crown detection [19].

More recently, YOLO-based segmentation models (e.g., YOLOv8/YOLOv11-seg) have emerged as promising one-stage alternatives. Their advantage lies in balancing detection accuracy with very fast inference speeds, enabling near real-time crown mapping from UAV flights. Studies report that lightweight YOLO variants, such as YOLOv8n, can process images efficiently on resource-constrained devices while still achieving competitive accuracy [20]. This makes YOLO particularly attractive for operational deployment across large orchards, where scalability and speed are crucial. However, these models often prioritize bounding-box localization and may lose precision in fine-grained boundary delineation compared with instance-focused methods. Thus, while YOLO offers practical benefits in speed and deployment readiness, its trade-off in segmentation granularity raises questions about its suitability for biovolume estimation without further refinement.

These differences highlight that no single model offers an all-purpose solution: Mask R-CNN excels in accuracy, U-Net in efficient semantic mapping, and YOLO in operational scalability. Comparative evaluation prior to deployment is therefore essential, as model choice must weigh accuracy, granularity, and computational feasibility to ensure robust biovolume estimation in precision agriculture.

## III. MATERIAL AND METHOD

### A. Study Area and Data Aquisition

The study site is an olive orchard located in Vicopisano, Tuscany, Italy (43o42'47" N 10o34'57" E) (Fig. 1). UAV imagery was acquired on 22 May 2022 at 9:46 AM under clear sky conditions, covering an area of 8568 m$^2$. A total of 333 UAV images were captured using a Parrot Anafi (version 1.8.2) drone equipped with a 21 MP RGB camera (1-inch CMOS sensor, with 4.0 mm focal length) was used to capture 333 images at constant altitude of 17 m above ground. The resulting ground sampling distance (GSD) is 0.55 cm/pixel, providing ultra- high-resolution detail of tee canopies and their shadows. Images were captured from various viewing angles (traverse and longitudinal) to ensure complete coverage of the orchard and to assist in photogrammetric reconstruction.

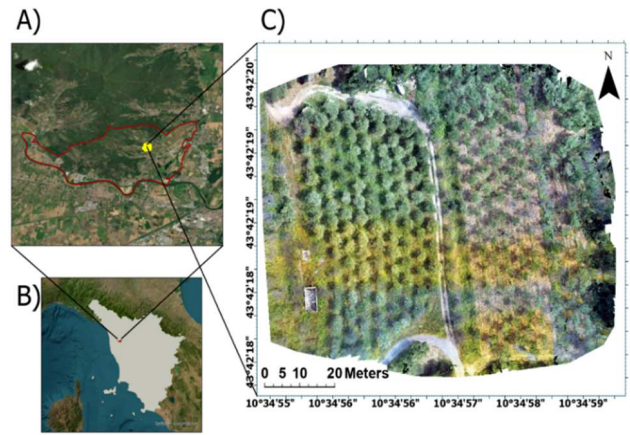

Fig. 1: Test location of image captured in Vicopisano, Tuscany in Italy (A, B), Digital orthophoto map of olive orchards (C).

### B. Data Pre-processing and Annotation

The 333 images (pixel size of 4608 x 3456) were processed using Agisoft Metashape to generate an orthomosaic and digital surface model (DSM) of the orchard.

The georeferenced orthomosaic (Fig. 2) provided a top-down view in which individual olive trees and their shadows are clearly distinguishable. Using this orthophoto, we semi-automatically annotated the outline of each olive tree crown and its corresponding shadow.

To generate reliable crown and shadow delineations, a hybrid workflow combining ArcGIS Pro preprocessing and CVAT labeling tool was adopted. In ArcGIS Pro, the raw RGB UAV imagery was transformed into HSV color space using the raster calculator with a conditional argument. This transformation emphasized brightness differences, enabling the systematic detection of shadow regions. By applying a 50% threshold on the value (V) channel, candidate shadow areas were identified, producing a control layer for subsequent segmentation.

In parallel, tree crown and shadow delineation was conducted in CVAT labeling tool, where the Segment Anything Model (SAM) provided a semi-automated labeling of individual tree objects. SAM's instance-level masks captured fine crown boundaries and their associated shadows with higher spatial precision. To ensure consistency and reduce over-segmentation, the HSV-derived shadow layer from ArcGIS Pro was used as a reference, effectively serving as a control to validate and refine the CVAT outputs.

This dual approach leveraged the spectral thresholding strength of ArcGIS for reliable shadow identification and the instance segmentation capability of SAM in CVAT for detailed object delineation. The integration of both ensured that tree crowns and their corresponding shadows were consistently and accurately mapped, thereby providing robust training data for subsequent biovolume estimation.

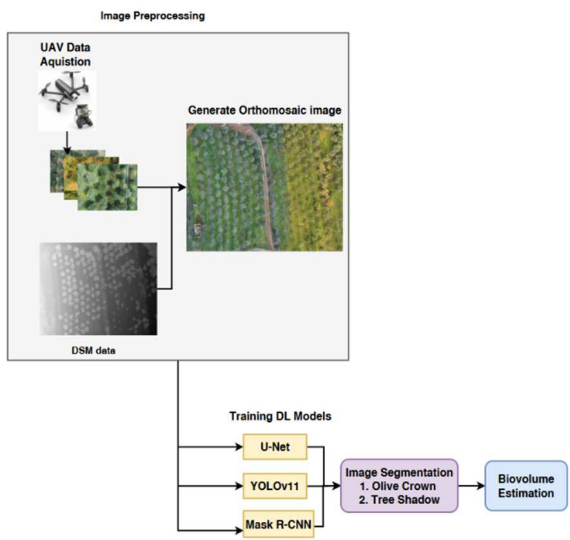

Fig. 2 Proposed method

### C. Deep Learning Models

We evaluated three deep learning segmentation models representing different state-of-the-art approaches: (1) U-Net (semantic segmentation), (2) YOLOv11m-seg (one-stage instance segmentation), and (3) Mask R-CNN (two-stage instance segmentation).

1. *U-Net*: a prominent encoder-decoder segmentation network, distinctively recognized by its "U" shape and symmetric encoder and decoder sections [21]. This architecture was developed to overcome some of the limitations of FCNs, particularly in scenarios with limited training data and a high demand for precise segmentation [21]. Originally designed for biomedical imaging with limited training data, U-Net excels in capturing fine-grained features and achieves high segmentation accuracy even with small datasets. U-Net has also widely used to semantically segment various UAV imagery with varying degrees of success. While it offers advantages such as simplicity and effectiveness in certain contexts, several limitations have been identified when using U-Net on UAV datasets, such as loss of spatial detail during down-sampling operations [22], and struggling to handle small and dynamic objects captured by UAVs [23].

2. *YOLOv11(You Only Look Once)*: is the latest model in the YOLO series, designed for real-time object detection and instance segmentation. It uses a one-stage detection architecture that enables rapid inference by predicting bounding boxes and class probabilities in a single pass [24]. YOLOv11 introduces refined feature extraction and optimized network designs to enhance speed, accuracy, and computational efficiency [24]. It supports multiple vision tasks and offers model variants from nano to extra-large, balancing performance and resource demands [24]. Though smaller models favor speed over peak accuracy, YOLOv11's high FPS and precision make it ideal for real-time applications like agricultural monitoring and automation [24].

3. *Mask R-CNN*: is a two-stage instance segmentation model that extends Faster R-CNN by adding a parallel mask prediction branch [25]. It uses a convolutional backbone (e.g., ResNet50) and a Region Proposal Network (RPN) to generate candidate regions (RoI). Features are extracted from each region using RoIAlign, which preserves spatial accuracy. These features are processed by three heads: classification, bounding-box regression, and a fully convolutional mask branch that outputs a binary mask for each instance, specific to its predicted class [25]. By predicting masks independently per RoI for each class, Mask R-CNN improves accuracy and enables simultaneous object detection and fine-grained instance segmentation with high precision [25].

All models were implemented in PyTorch environment and trained on an NVIDIA Tesla V100 GPU (32 GB). The training duration was capped at 250 epochs, with early stopping triggered if validation loss failed to improve for 25 consecutive epochs. Learning rates followed a cosine annealing schedule with a minimum reduction factor of 0.1, and mixed-precision training (AMP) was enabled to optimize GPU usage

### D. Image Segmentation

The core of the proposed method involves the segmentation of two key elements from the UAV imagery: olive tree crowns and their respective shadows.

1. *Olive crown segmentation*: This process precisely delineates individual tree crowns from UAV orthoimages. DL models (U-Net, YOLOv11, Mask R-CNN) trained on annotated data to identify and segment crown pixels, producing binary masks for each tree. These masks directly provide the crown projected area (CPA), vital input for biovolume estimation. Challenges may include differentiating overlapping crowns and handling visual variability, yet accurate CPA is crucial for agricultural assessment.

2. *Tree shadow segmentation:* This process identifies and delineates shadows cast by olive trees in UAV orthoimages.

DL models are trained to segment pixels corresponding to these tree-specific shadows, outputting binary masks. The primary significance lies in using the segmented shadow length, combined with GSD and solar geometry, to accurately estimate tree height. Challenges include differentiating tree shadows from other objects and managing varied surface conditions under the shadows.

*E. Model Evaluation*

To evaluate the performance of different models, we use metrics such as precision, recall, and F1-score, along with mean Intersection-over-Union (mIoU). These metrics provide complementary insights into segmentation performance, where precision and recall highlight error types, F1 balances both, and mIoU capture boundary quality.

*F. Biovolume Estimation*

The biovolume (V) of individual olive trees will be estimated as a proxy using the segmented crown area and the derived tree height. This approach typically models the tree canopy as a simple geometric solid (e.g., cylinder, ellipsoid, or a general volumetric proxy). This study uses the common simplified general volumetric proxy as shown in Fig. 3.

$$V \approx A_c \times H_t \tag{1}$$

Where:
- $A_c$ (Crown Projected Area): Calculated from the segmented crown mask. If $N_{crown\ pixels}$ is the number of pixels in a segmented crown mask, and $GSD_x$ and $GSD_y$ are the ground sampling distances in x and y dimensions respectively, then:

$$A_c = N_{crown\ pixels} \times (GSD_x \times GSD_y) \tag{2}$$

- $H_t$ (Tree Height): Estimated from the length of the segmented tree shadow ($L_s$) and the solar elevation angle ($\alpha$) or solar zenith angle ($\theta_z = 90° - \alpha$) at the time and location of image acquisition. Solar angles will be calculated using established astronomical algorithms based on date, time, and geographic coordinates using astonomical algorithm (e.g., https://www.suncalc.org ).

$$H_t = L_s / \tan(\alpha) = L_s \times \tan(\theta_z) \tag{3}$$

- $L_s$ (Shadow Length) is measured from the base of the tree to the tip of its shadow in the direction opposite to the sun's azimuth minus 0.8 m (obtained from field visit) to exclude the lower, unbranched tree part. Accurate determination requires proper identification of the shadow's extent.

*G. Assumptions*
- The ground surface beneath the tree and its shadow is flat and level with the tree base. For undulating terrain, a Digital Elevation Model (DEM) or Digital Surface Model (DSM) required for more accurate height calculations, which is beyond the primary scope of this specific comparison but a noted consideration.
- The estimated biovolume is a proxy value and does not account for complex crown porosity or internal canopy structure. More sophisticated allometric equations or form factors (k, in $V = k \times A_c \times H_{it}$) could be applied if specific relationships for olive trees in the study area are known or can be derived.
- Shadows are distinct and primarily cast by the tree crown.

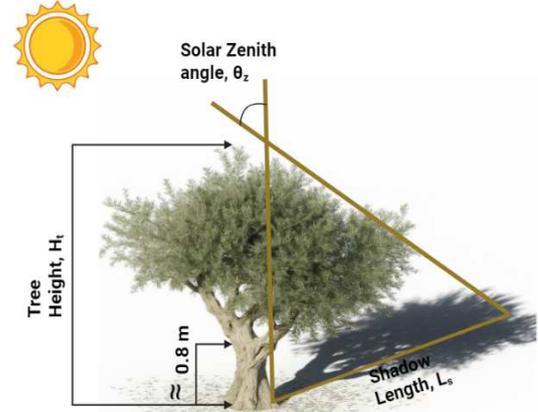

Fig. 3 Illustration of olive tree biovolume estimation using DL segmentation models to automatically extract tree crown and shadow polygon using UAV RGB images. Tree canopy volume is calculated by multiplying Ac with Ht. Olive tree image by macro-vector (https://freepik.com, accessed 05 June 2025).

## IV. EXPERIMENTAL RESULT

In this section, we analyze the details of training models, including the data set used, setting up deep learning models, and evaluating the results of different methods.

*A. Dataset and Preparation*

UAV images with an original resolution of 4608 × 3456 pixels were resized to 1500 × 1500 to standardize processing while retaining canopy details. To mitigate sampling bias, the dataset was split using the Monte Carlo method, allocating 70% for training, 20% for validation, and 10% for testing. This ensured a robust and balanced evaluation across olive crown and shadow classes.

*B. Model Training Protocol*

Three state-of-the-art deep learning models were trained under comparable experimental settings:

The U-Net model was optimized using the AdamW algorithm with an initial learning rate of $3\times10^{-4}$ and weight decay of $5\times10^{-4}$. A hybrid loss function combining cross-entropy and class-weighted Dice loss was employed to address the imbalance between tree crowns, shadows, and background pixels. Class weights were estimated from the inverse log-frequency distribution of training masks, and shadow-containing tiles were oversampled to further improve representation of the minority class.

YOLOv11m-seg was initialized with pretrained weights (~22M parameters) and trained with AdamW under the same learning rate and weight decay settings as U-Net. Its composite objective integrated classification, bounding-box regression based on mIoU,

Mask R-CNN with a ResNet50-FPN-v2 backbone was trained with AdamW using identical optimization parameters as YOLOv11m-seg and U-Net. Its multi-task loss consisted of bounding boxes, and binary cross-entropy for masks, with equal weighting between detection and segmentation objectives to enhance pixel-level accuracy.

C. Quantitative Performance

Table 1 presents the comparison of segmentation models in terms of the number of layers, parameters, and inference and training time.

Table 1 Comparing segmentation models: layers parameters, training and inference times

| Models | No. of Layers | No. of Parameters (M) | Training Time (h) | Inference Time (s) |
|---|---|---|---|---|
| U-Net | 41 | 31 | 1.42 | 0.42 |
| YOLOv11m-seg | 268 | 22 | 0.98 | 0.12 |
| Mask R-CNN | 235 | 44 | 1.65 | 0.35 |

Among the three tested architectures, YOLOv11m-seg demonstrates a clear advantage in terms of computational efficiency. With 268 layers and 22 million parameters, it required only 0.98 hours of training time and achieved the fastest inference speed of 0.12 seconds per image. In contrast, U-Net (23 layers, 31M parameters) and Mask R-CNN (235 layers, 44M parameters) were notably heavier, with training times of 1.42 hours and 1.65 hours, and inference times of 0.42 seconds and 0.35 seconds, respectively.

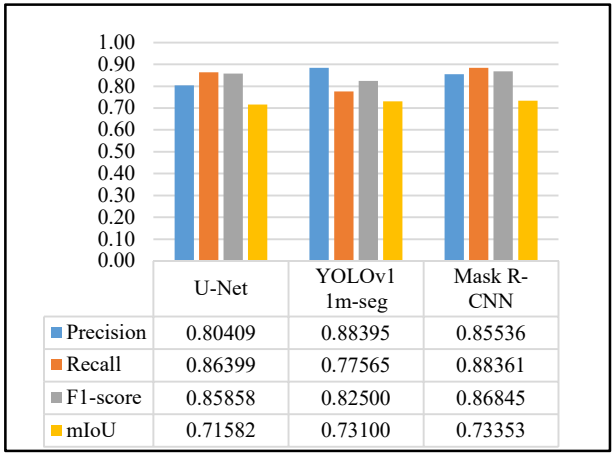

Fig 4 Performance metrics of the three segmentation models

The comparative analysis highlights distinctive strengths and weaknesses among U-Net, YOLOv11m-seg, and Mask R-CNN (Fig 4). The result demonestrate clear performance trade-offs among the three models. U-Net achieved lowest preision (0.804) but relatively high recall (0.864), indicating strong sensitivity to crown and shadow detection but a tendency to include background noise. Its F1-score (0.859) and lower mIoU (0.716) highlights the overlaping issues. This suggests U-Net is suitable when sensitivity is prioritized over accuracy in boundary delination.

YOLOv11m-seg recorded the highest precision (0.884), excelling at reducing false positives and producing clean segmentations. However, its recall (0.776) was weakest, revealing a risk of under-segmentation, especially in dense or shaded canopies. The resulting F1-score (0.825) and mIoU (0.731) were moderate, suggesting YOLOv11m-seg is more reliable for large-scale, real-time applications where efficiency and precision outweigh completeness.

Mask R-CNN outperformed both alternatives in recall (0.884), F1-score (0.868), and mIoU (0.734), demonstrating its ability to capture canopy extent while maintaining accuracy. Although computationally heavier, its superior consistency across metrics makes it the most robust choice for precise biovolume estimation and research-focused applications.

The visual comparison (Fig 5) further illustrates these trade-offs. Mask R-CNN delineated crowns with clear and consistent boundaries, even in dense orchards, aligning with its top performance metrics. YOLOv11m-seg produced smooth, precise crown contours but under-segmented smaller or partially occluding trees, consistent with its lower recall. U-Net generated broad crown coverage but also extended into background regions, producing noisy masks that match its lower precision.

These results suggest Mask R-CNN is most suitable for research and precise biovolume estimation, YOLOv11m-seg balances accuracy with efficiency for large-scale orchard monitoring, while U-Net offers a lightweight option where computational simplicity and high sensitivity are prioritized.

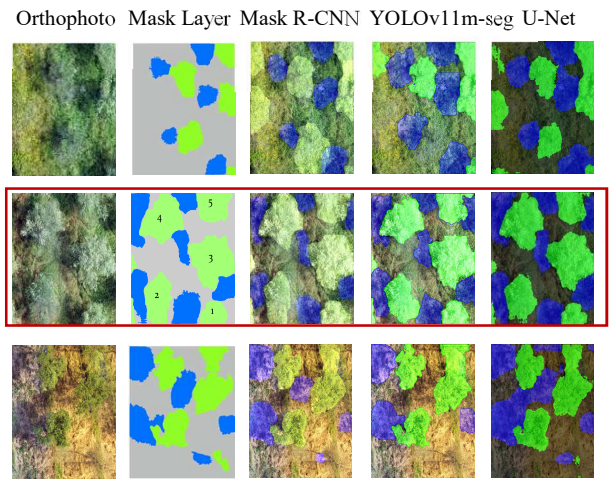

Fig 5 Comparison of olive tree crown and shadow segmentation: orthophoto, ground-truth mask, Mask R-CNN, YOLOv11m-seg, and U-Net outputs. The red bounded images are taken for sample calculation of

D. Result of Tree Biovolume Estimation

Using Mask R-CNN segmentations on five sample trees (those red highlighted in Fig 5), we computed crown areas, shadow lengths, heights, and resulting biovolumes (Table 2). The estimated crown areas ranged from 2.33 to 8.90 m², tree heights from 1.83 to 3.14 m, and biovolumes from 4.36 to 24.35 m³. Larger crowns with longer shadows produced significantly higher volumes. This confirms that combining segmented canopy area with shadow-derived height yields a plausible structural metric. For instance, Tree 4 (crown area 7.75 m², height 3.14 m) had the highest volume (24.35 m³). These results demonstrate the method's ability to reflect relative tree size differences in the orchard.

Table 2 Estimated tree biovolumes for sample trees (IDs as in Fig 5). Crown area from Mask R-CNN segmentation, shadow length measured from orthophoto, height from solar geometry, and volume = $A_c \times H_t$

| Tree ID | Crown Area (m$^2$) | Shadow Length (m) | Height (m) | Biovolume (m$^3$) |
|---|---|---|---|---|
| 1 | 2.326 | 2.326 | 1.874 | 4.360 |
| 2 | 5.149 | 3.382 | 3.088 | 15.902 |
| 3 | 8.895 | 2.326 | 1.874 | 16.673 |
| 4 | 7.754 | 3.428 | 3.140 | 24.349 |
| 5 | 3.856 | 2.288 | 1.830 | 7.057 |

Overall, errors in segmentation would directly affect volume. Under-segmentation (missed canopy) would underestimate volume, while over-segmentation inflates it. In our analysis, Mask R-CNN's high accuracy suggests the biovolume estimates are reliable proxies. The main limitation is lack of ground-truth field volumes for validation. Future work will integrate in-situ measurements, high-resolution DEM/DSM data, and species-specific allometric models to calibrate and validate these biovolume estimates.

## V. CONCLUSION

This study systematically compared U-Net, YOLOv11m-seg, and Mask R-CNN for segmenting olive tree crowns and shadows from UAV imagery, with the goal of estimating individual tree biovolume. Mask R-CNN consistently outperformed the others in accuracy (highest F1-score and mIoU), indicating its superior capability for detailed canopy delineation. YOLOv11m-seg provided a favorable speed–accuracy trade-off, excelling in inference speed and precision, while U-Net was fastest to train and highly sensitive but less precise. Integrating the segmented crown area with shadow-based height proved effective for volume approximation, confirming that combined spatial features can serve as a structural proxy.

In practical terms, Mask R-CNN is recommended when accuracy is paramount, whereas YOLOv11m-seg may be preferred for rapid monitoring of large orchards. U-Net remains an option when computational resources are limited, given its simpler architecture. Ultimately, deploying such models can greatly enhance precision olive farming by enabling automated tree-by-tree analytics.

Future work will validate these estimates against field data and refine the method. Specifically, incorporating in-situ crown and height measurements, linking to LiDAR/DEM terrain data, and applying olive-specific growth models will improve quantitative biovolume accuracy. This will ensure robust, generalizable deep-learning tools for precision agriculture in changing climates.